# The trimer-based spin liquid candidate Ba$_4$NbIr$_3$O$_{12}$

Loi T. Nguyen and R.J. Cava

Department of Chemistry, Princeton University, Princeton, New Jersey 08544, USA

**Abstract**

Ba$_4$NbIr$_3$O$_{12}$, a previously unreported material with a triangular planar geometry of Ir$_3$O$_{12}$ trimers, is described. Magnetic susceptibility measurements show no magnetic ordering down to 1.8 K despite the Curie-Weiss temperature of -13 K. The material has a very low effective magnetic moment of 0.80 µ$_B$/f.u. To look at the lower temperature behavior, the specific heat (C$_p$) was measured down to 0.35 K; it shows no indication of magnetic ordering and fitting a power law to C$_p$ vs. T below 2 K yields the power α = ¾. Comparison to the previously unreported trimer compound made with the 4$d$ element Rh in place of the 5$d$ element Ir, Ba$_4$NbRh$_3$O$_{12}$, is presented. The analysis suggests that Ba$_4$NbIr$_3$O$_{12}$ is a candidate spin liquid material.

**Introduction**

Spin orbit coupling (SOC), the interaction between the spin of an electron and the magnetic field present in its orbit around a heavy nucleus, has proven to be an important factor for understanding the magnetic properties and insulating state in many iridium-based oxides.[1],[2],[3],[4] In an octahedral crystal field, in the absence of SOC, the *d*-orbitals split into triply degenerate $t_{2g}$ and doubly degenerate $e_g$ states, but SOC can further split the $t_{2g}$ manifold into two discrete energy levels, quartet J = 3/2 and doublet J = ½. The importance of SOC for understanding the magnetism in Ir-based compounds therefore depends on the number of electrons in the 5*d* manifold of electronic states. Here we consider a system with two formally $Ir^{4+}$ ($5d^5$) and one formally $Ir^{3+}$ ($5d^6$) present. Spin liquid behavior, the spin Hall effect[5], the topological Weyl semi-metallic state, and Mott insulator states[3],[6] are all present in iridates, which have also been of recent interest because spin liquid states may be stable at low temperatures in some of these materials.[8] [9] [10] Thus new materials in this family are of significant interest.

Here we describe a previously unreported Ir-trimer based compound, $Ba_4NbIr_3O_{12}$, which displays a rhombohedral structure with the space group *R*-3*m* (No. 166). The rhombohedral symmetry displayed is in contrast to the monoclinic symmetry (*C*2/*m*) found for materials in the related $Ba_4LnIr_3O_{12}$ (Ln=lanthanide) family. In addition, we describe previously unreported $Ba_4NbRh_3O_{12}$, which forms $Rh_3O_{12}$ trimers. The comparison is of interest due to the fact that the materials are based on magnetically active 5*d* (Ir) and 4*d* (Rh) elements with the same number of valence electrons. Neither compound shows any magnetic ordering down to 1.8 K even though their antiferromagnetic Curie-Weiss temperatures are larger than 1.8 K, -13 K for $Ba_4NbIr_3O_{12}$ and -23 K for $Ba_4NbRh_3O_{12}$. Magnetic measurements reveal an effective moment of 0.80 μ$_B$/f.u for $Ba_4NbIr_3O_{12}$ and 1.48 μ$_B$/f.u for $Ba_4NbRh_3O_{12}$. Resistivity measurements show that both compounds are semiconductors with transport gaps of about 0.05 eV. $Ba_4NbRh_3O_{12}$ shows a feature in its specific heat that can be interpreted as being the result of magnetic ordering at 1.5 K. $Ba_4NbIr_3O_{12}$ on the other hand displays a linear upturn in $C_p/T$ at low temperatures and therefore appears to be a candidate for a spin liquid.

**Experimental**

Polycrystalline samples of $Ba_4NbIr_3O_{12}$ and $Ba_4NbRh_3O_{12}$ were synthesized by solid-state reaction using $BaCO_3$, $Nb_2O_5$, $IrO_2$ and $RhO_2$ (Alfa Aesar, 99.9%, 99.5%, 99.9% and 99.9% respectively) in stoichiometric ratios as starting materials. Reagents were mixed thoroughly, placed in alumina crucibles, and heated in air at 900 °C for 24 hours. The resulting powders were re-ground, pressed into pellets and heated in air at 1100 °C for 48 hours. The phase purities and crystal structures were determined through powder X-ray diffraction (PXRD) using a Bruker D8 Advance Eco with Cu Kα radiation and a LynxEye-XE detector. The structural refinements were performed with *GSAS*[11]. The crystal structure drawings were created by using the program *VESTA*[12].

The magnetic susceptibilities of $Ba_4NbM_3O_{12}$ (M = Ir or Rh) powders were measured in a Quantum Design Physical Property Measurement System (PPMS) DynaCool equipped with a VSM option. The magnetic susceptibilities between 1.8 and 300 K, defined as *M/H*, where *M* is the sample magnetization and *H* is the applied field, were measured at the field of *H* = 5 kOe for $Ba_4NbIr_3O_{12}$ and 2 kOe for $Ba_4NbRh_3O_{12}$; some additional measurements were performed in an applied field of 100 Oe. The resistivities were measured by the DC four-contact method in the temperature range 200 K to 300 K in the PPMS. For the resistivity measurements, the samples were pressed, sintered, and cut into pieces with the approximate size $1.0 \times 2.0 \times 1.0$ mm³. Four Pt contact wires were connected to the samples using silver paint. The specific heat was measured from 20 K to 1.8 K by a PPMS DynaCool equipped with a heat capacity option, and by using a $^3$He system to reach down to 0.35 K.

**Results and discussion**

The powder X-ray diffraction patterns and structural refinements for $Ba_4NbIr_3O_{12}$ and $Ba_4NbRh_3O_{12}$ are shown in **Figures 1** and **2**. The structure of $Ba_4NbRu_3O_{12}$[13] was used as the starting model for the structure refinements. The agreement between the observed and calculated patterns is excellent. Both $Ba_4NbIr_3O_{12}$ and $Ba_4NbRh_3O_{12}$ are rhombohedral, with the space group *R-3m* (No. 166). The lattice parameters and structural parameters for both compounds are summarized in **Table 1**. Their structure consists of three $MO_6$ (M = Ir or Rh) octahedra connected by face-sharing to form $M_3O_{12}$ trimers. The trimers are arranged in a triangular planar lattice. While $Ba_4LnIr_3O_{12}$ (Ln = Lanthanides) have the monoclinic *C2/m* space group, $Ba_4NbIr_3O_{12}$ adopts a higher symmetry *R-3m* crystal structure[14][15]. The individual $NbO_6$ octahedra and $M_3O_{12}$ trimers in these $Ba_4NbM_3O_{12}$ phases alternate along *c* to generate the 12-layer (i.e. three layers of 4 octahedra) hexagonal perovskite structure. Individual $M_3O_{12}$ trimers are corner-sharing with the non-magnetic $NbO_6$ octahedra, and not to other trimers, such that the magnetic coupling between trimers is of the M-O-O-M super-super exchange type. Structure refinements in which the Ir or Rh to Nb ratio was allowed to vary were highly unsatisfactory, thus confirming the composition of the materials. To the resolution of the structural determination, the Nb:Ir and Nb:Rh sublattices are fully ordered.

The temperature-dependent magnetic susceptibility of $Ba_4NbIr_3O_{12}$ and its reciprocal are plotted in **Figure 3a**. The data from 100 K to 300 K are well fit to the Curie-Weiss law $\chi = \frac{C}{T-\theta_{CW}} + \chi_o$, where $\chi_o$ is the temperature-independent part of the susceptibility, *C* is the Curie constant, and $\theta_{CW}$ is the Curie-Weiss temperature. The least squares fitting yields $\chi_o$ = -1.75x10$^-$

$10^{-4}$ emu Oe$^{-1}$ mol-f.u.$^{-1}$, $\mu_{eff}$ = 0.80 $\mu_B$/f.u, and $\theta_{CW}$ = -13 K. Similar magnetic behavior was observed in Ba$_5$AlIr$_2$O$_{11}$[16], which has a Curie-Weiss temperature of -14 K and effective moment of 1.04 $\mu_B$/f.u. The effective moment of Ba$_4$NbIr$_3$O$_{12}$ is smaller than the one observed in Ba$_4$Ni$_{1.94}$Ir$_{2.06}$O$_{12}$, which has a mixed trimer between Ir and Ni.[17] Similar characterization of Ba$_4$NbRh$_3$O$_{12}$ gives $\chi_o$ = 2.1x10$^{-4}$ emu Oe$^{-1}$ mol-f.u.$^{-1}$, $\mu_{eff}$ = 1.48 $\mu_B$/f.u, and $\theta_{CW}$ = -23 K, as illustrated in **Figure 3b**. Ir and Rh are in the same column of the periodic table, Rh being a 4$d$ element and Ir being a 5$d$ element, and explanation of the differences in effective moments in the materials may be of future interest. The temperature-dependent magnetic susceptibilities of Ba$_4$NbIr$_3$O$_{12}$ and Ba$_4$NbRh$_3$O$_{12}$ under different applied fields are shown in the Supplementary Information, **Figures S1**.

Different from the case of Ba$_4$Ni$_{1.94}$Ir$_{2.06}$O$_{12}$[15] where the magnetization comes from both the Ir$_2$NiO$_{12}$ trimers and the isolated Ni$^{2+}$O$_6$ octahedra, in Ba$_4$NbM$_3$O$_{12}$, where both Ba$^{2+}$ and Nb$^{5+}$ are non-magnetic, the magnetic properties of the Ba$_4$NbM$_3$O$_{12}$ materials studied here are determined by the intertrimer and intratrimer interactions of the M$_3$O$_{12}$ units; we postulate that each trimer acts as an electronic and magnetic building block in this material, with M2-M1-M2 bonding interactions within the trimers. The resulting antiferromagnetic interactions lead to $\theta_{CW}$ = -13 K for the Ir case and -23 K for the Rh case.

**Figures 4a,b** show the field-cooled (FC) and zero-field-cooled (ZFC) DC susceptibility in an applied field of 100 Oe for Ba$_4$NbIr$_3$O$_{12}$ and Ba$_4$NbRh$_3$O$_{12}$. In both compounds, the magnetic susceptibility increases down to the lowest measured temperature of 1.8 K. Combined with relatively large negative Curie-Weiss temperature, the behavior in the ZFC/FC DC susceptibility indicates the possibility of magnetic frustration (frustration index = $|\theta_{CW}|/T_M \approx 7$) or even spin liquid behavior in these materials. **Figures 4c,d** show the magnetizations of Ba$_4$NbM$_3$O$_{12}$ as a function of applied field up to $\mu_o H$ = 9 T at different temperature from 2 to 100 K. The magnetization is directly proportional to the applied field up to fields well above the fields used for the chi vs. T measurements (0.2 and 0.5 Tesla). Although M vs H is curved at higher fields, there is no sign of saturation up to the highest field applied.

**Figure 5** shows the specific heat divided by temperature for $Ba_4NbIr_3O_{12}$ and $Ba_4NbRh_3O_{12}$ in the temperature range from 1.8 K to 200 K. There is no anomaly, indicating the absence of long-range magnetic order or phase transitions in this temperature range. We note that at 200 K, $C_p$ has not yet reached the saturation value of 3NR (N is the number of atoms), but this behavior observed is often encountered in materials where different atomic masses and strong bonds between atoms lead to very high vibrational frequencies[19]. The upturn at the lowest temperature of this measurement motivated our heat capacity measurements under different fields. $C_p/T$ data for $Ba_4NbIr_3O_{12}$ and $Ba_4NbRh_3O_{12}$ from 1.8-20 K under the fields of 0, 1, 5 and 9 T are shown in **Figure S2**. The upturn in the iridate is suppressed by the applied field and there is no upturn seen at an applied field of 9 T. This behavior has recently observed in materials proposed to be quantum spin liquids.[20]-[24]

**Figure 6** shows the heat capacity divided by temperature for both materials measured down to 0.35 K. While $Ba_4NbRh_3O_{12}$ shows a broad anomaly at 1.5 K (**Figure 6c**) which we suggest is the signature of magnetic ordering, $Ba_4NbIr_3O_{12}$ has a linear upturn in $C_p/T$ below 2K (**Figure 6a**).[10] While the extrapolation of heat capacity goes to 0 in the case of $Ba_4NbRh_3O_{12}$ (**Figure 6d**), the heat capacity extrapolates to 15 mJ/mol-K at absolute zero temperature in $Ba_4NbIr_3O_{12}$ (**Figure 6c**). This suggests that $Ba_4NbIr_3O_{12}$ is a spin liquid candidate since the presence of a significant amount of magnetic entropy at low temperatures implies the presence of low energy spin excitations.[20]-[24] The heat capacity data for $Ba_4NbIr_3O_{12}$ below 2 K fits to the function $C_p(T) = kT^{\alpha}$, where k = 0.1 and $\alpha \approx$ ¾, as shown in **Figure 6b**; the power smaller than 1 suggests a possible spin liquid state in this material.

The resistivities of $Ba_4NbIr_3O_{12}$ and $Ba_4NbRh_3O_{12}$ are plotted as a function of reciprocal temperature in **Figure 7**. Resistivity data from 200 to 300 K were fit to the standard model $\rho = \rho_o e^{\frac{E_a}{k_bT}}$, and the transport activation energy $E_a$ was calculated to be approximately 0.05 eV for both materials. With the activation energy of 0.05 eV, $Ba_4NbIr_3O_{12}$ and $Ba_4NbRh_3O_{12}$ are both semiconductors, similar to the related trimer-based compounds $Ba_4NbRu_3O_{12}$, $Ba_4LnRu_3O_{12}$ and $Ba_4LnIr_3O_{12}$.[13],[25]

**Conclusion**

$Ba_4NbIr_3O_{12}$ and $Ba_4NbRh_3O_{12}$ crystallize in a 12-layer hexagonal perovskite structure in the *R*-3*m* space group; $Ir_3O_{12}$ or $Rh_3O_{12}$ trimers are the magnetically active parts of the structure because $Ba^{2+}$ and $Nb^{5+}$ are not magnetic. While $Ba_4NbRh_3O_{12}$ appears to show magnetic ordering at 1.5 K, the low temperature heat capacity of $Ba_4NbIr_3O_{12}$ follows a power law between 2 and 0.35 K, implying that the material is a candidate for a spin liquid. First principles calculations will be of interest to investigate the electronic structure of this material. Very low temperature thermodynamic and neutron diffraction characterization will be of interest to confirm the presence or absence of a spin liquid state.

**Acknowledgement**

This work was supported as part of the Institute for Quantum Matter, an Energy Frontier Research Center funded by the U.S. Department of Energy, Office of Science, Office of Basic Energy Sciences under Award Number DE-SC0019331.

**Table 1.** Structural parameters for $Ba_4NbIr_3O_{12}$ and $Ba_4NbRh_3O_{12}$ at 300 K. Both compounds crystallize in the *R-3m* space group (No. 166).

| Atom | Wyckoff. | Occ. | x | y | z | $U_{iso}$ |
|---|---|---|---|---|---|---|
| $Ba_4NbIr_3O_{12}$ | | | | | | |
| Ba1 | 6c | 1 | 0 | 0 | 0.12890(6) | 0.0302(7) |
| Ba2 | 6c | 1 | 0 | 0 | 0.28585(7) | 0.0326(8) |
| Nb | 3a | 1 | 0 | 0 | 0 | 0.0008(4) |
| Ir1 | 3b | 1 | 0 | 0 | ½ | 0.0298(4) |
| Ir2 | 6c | 1 | 0 | 0 | 0.41148(6) | 0.0386(7) |
| O1 | 18h | 1 | 0.4817(7) | 0.5183(7) | 0.1216(3) | 0.067(5) |
| O2 | 18h | 1 | 0.5146(6) | 0.4854(6) | 0.2956(4) | 0.066(4) |
| a = 5.7827(2) Å, c = 28.7725(9) Å | | | | | | |
| $\chi^2$ =1.38, $R_{wp}$ = 5.79%, $R_p$ = 4.48%, $R_F^2$ = 6.09% | | | | | | |
| $Ba_4NbRh_3O_{12}$ | | | | | | |
| Ba1 | 6c | 1 | 0 | 0 | 0.12843(9) | 0.0229(9) |
| Ba2 | 6c | 1 | 0 | 0 | 0.28570(9) | 0.0224(9) |
| Nb | 3a | 1 | 0 | 0 | 0 | 0.0090(3) |
| Rh1 | 3b | 1 | 0 | 0 | ½ | 0.0115(8) |
| Rh2 | 6c | 1 | 0 | 0 | 0.4106(1) | 0.0115(8) |
| O1 | 18h | 1 | 0.4848(4) | 0.5152(8) | 0.1202(4) | 0.002(4) |
| O2 | 18h | 1 | 0.510(1) | 0.490(1) | 0.2911(7) | 0.077(6) |
| a = 5.75707(4) Å, c = 28.4734(5) Å | | | | | | |
| $\chi^2$ =2.47, $R_{wp}$ = 5.89%, $R_p$ = 4.44%, $R_F^2$ = 9.57% | | | | | | |

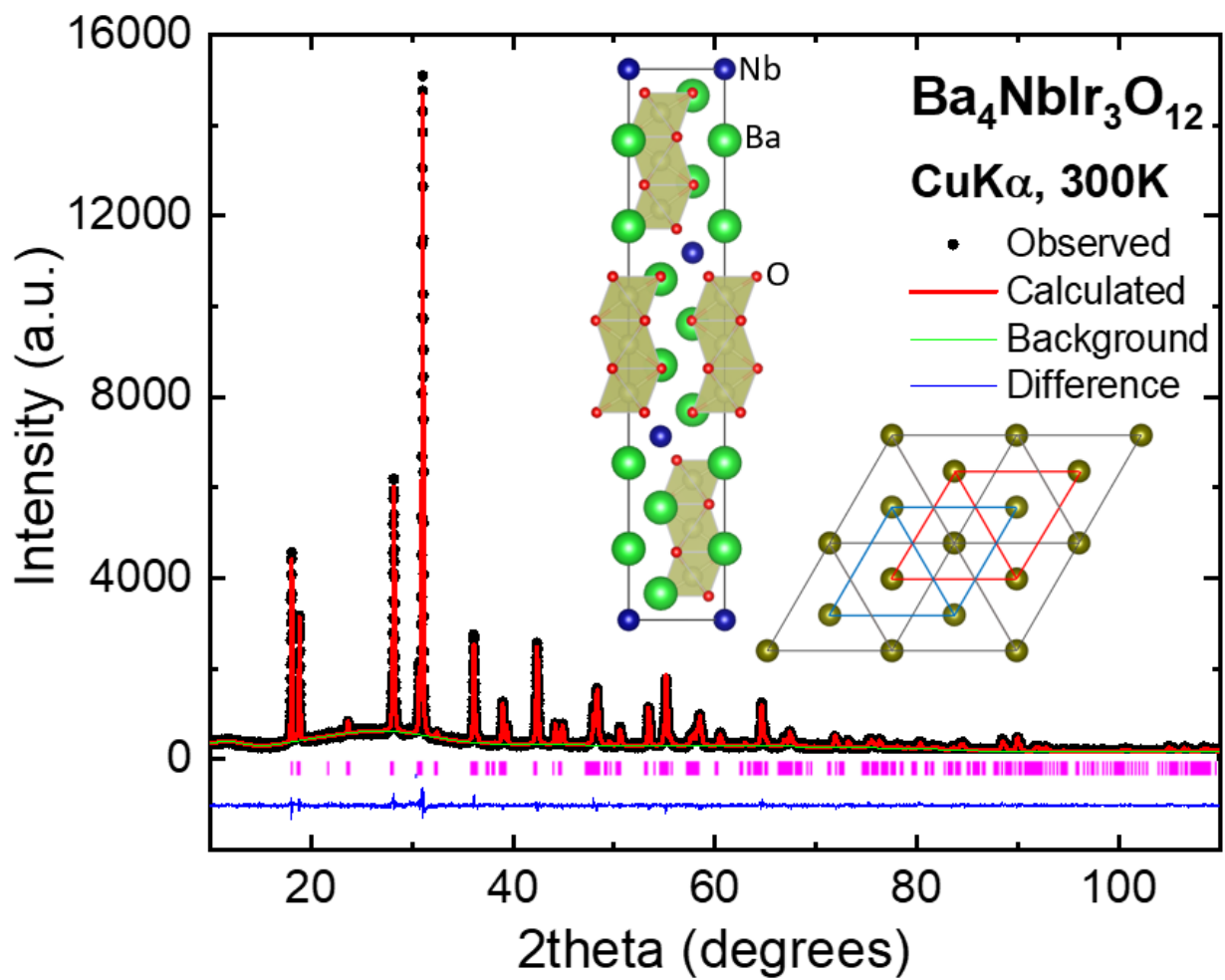

Figure 1: (Color online) Rietveld Powder X-ray diffraction refinement for $Ba_4NbIr_3O_{12}$ in space group *R-3m*. The observed X-ray pattern is shown in black, calculated in red, difference ($I_{obs}$-$I_{calc}$) in blue, and the tick marks denote allowed peak positions in pink. $R_p$ = 0.0448, $R_{wp}$ = 0.0579, $\chi^2$ = 1.38. The left insert shows the trimer crystal structure ($Ir_3O_{12}$ trimers are shaded yellow) and (right insert) that the trimers are arranged in an ABC packing array.

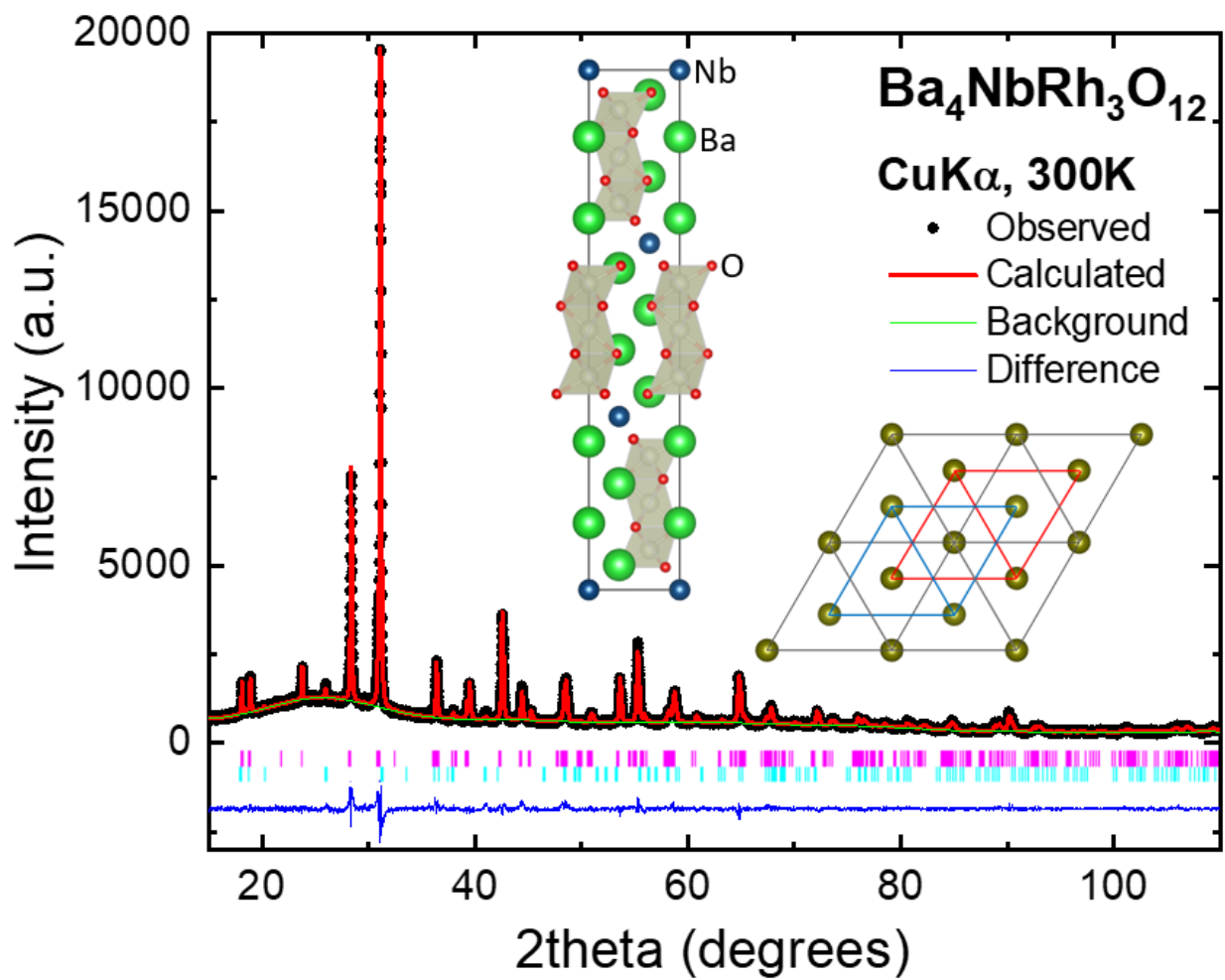

Figure 2: (Color online) Rietveld Powder X-ray diffraction refinement for $Ba_4NbRh_3O_{12}$ in space group *R*-3*m*. The observed X-ray pattern is shown in black, calculated in red, difference ($I_{obs}$-$I_{calc}$) in blue, and the tick marks denote allowed peak positions in pink ($Ba_4NbRh_3O_{12}$) and in cyan ($BaRhO_3$). $R_p$ = 0.0439, $R_{wp}$ = 0.0578, $\chi^2$ = 2.47. The left insert shows the trimer crystal structure ($Rh_3O_{12}$ trimers are shaded gray) and (right insert) that the trimers are arranged in an ABC packing array.

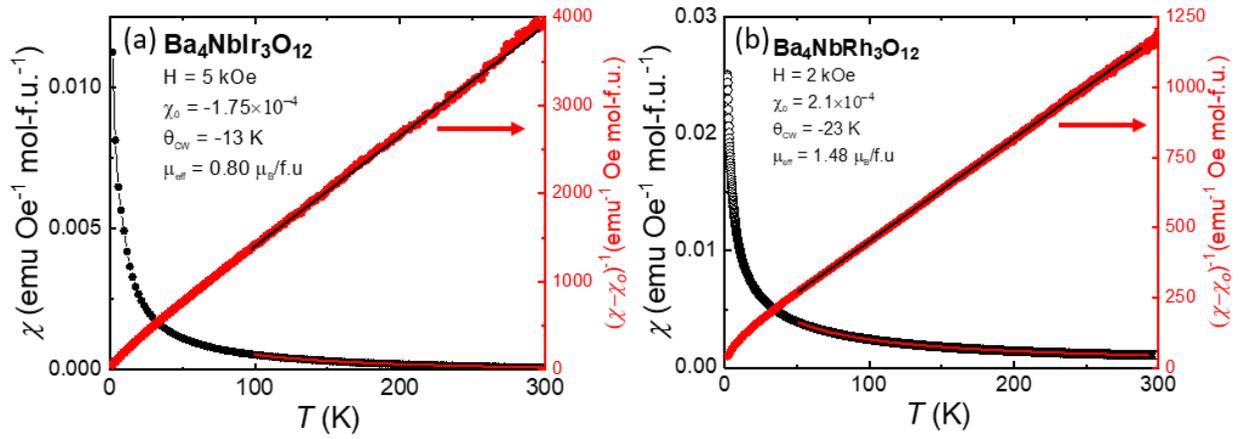

Figure 3: (Color online) The temperature dependence of the magnetic susceptibility and the inverse of the difference between the magnetic susceptibility and the temperature independent magnetic susceptibility ($\chi_0$) for (a) $Ba_4NbIr_3O_{12}$ and (b) $Ba_4NbRh_3O_{12}$. The red solid lines are the fits to the Curie-Weiss law.

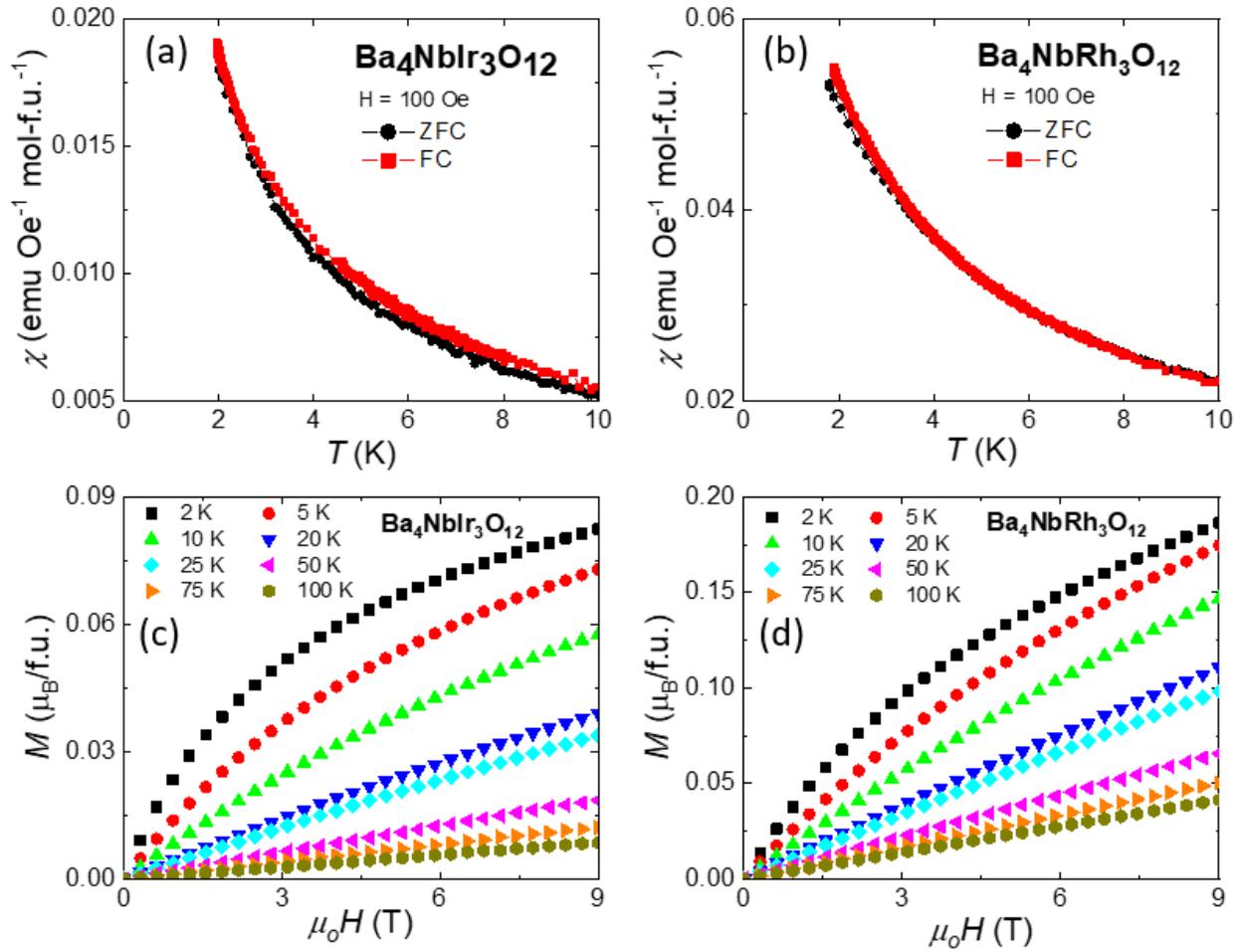

Figure 4: (Color online) Field Cooled (FC) and Zero Field Cooled (ZFC) DC magnetic susceptibility in an applied field of 100 Oe for $Ba_4NbIr_3O_{12}$ (a) and $Ba_4NbRh_3O_{12}$ (b) from 1.8-10 K. There is no bifurcation between FC and ZFC DC magnetic susceptibility down to 1.8 K in the former material and a small amount below 2 K in the latter. Magnetization of $Ba_4NbIr_3O_{12}$ (c) and $Ba_4NbRh_3O_{12}$ (d) as function of applied filed from 0-9 T, at various temperatures.

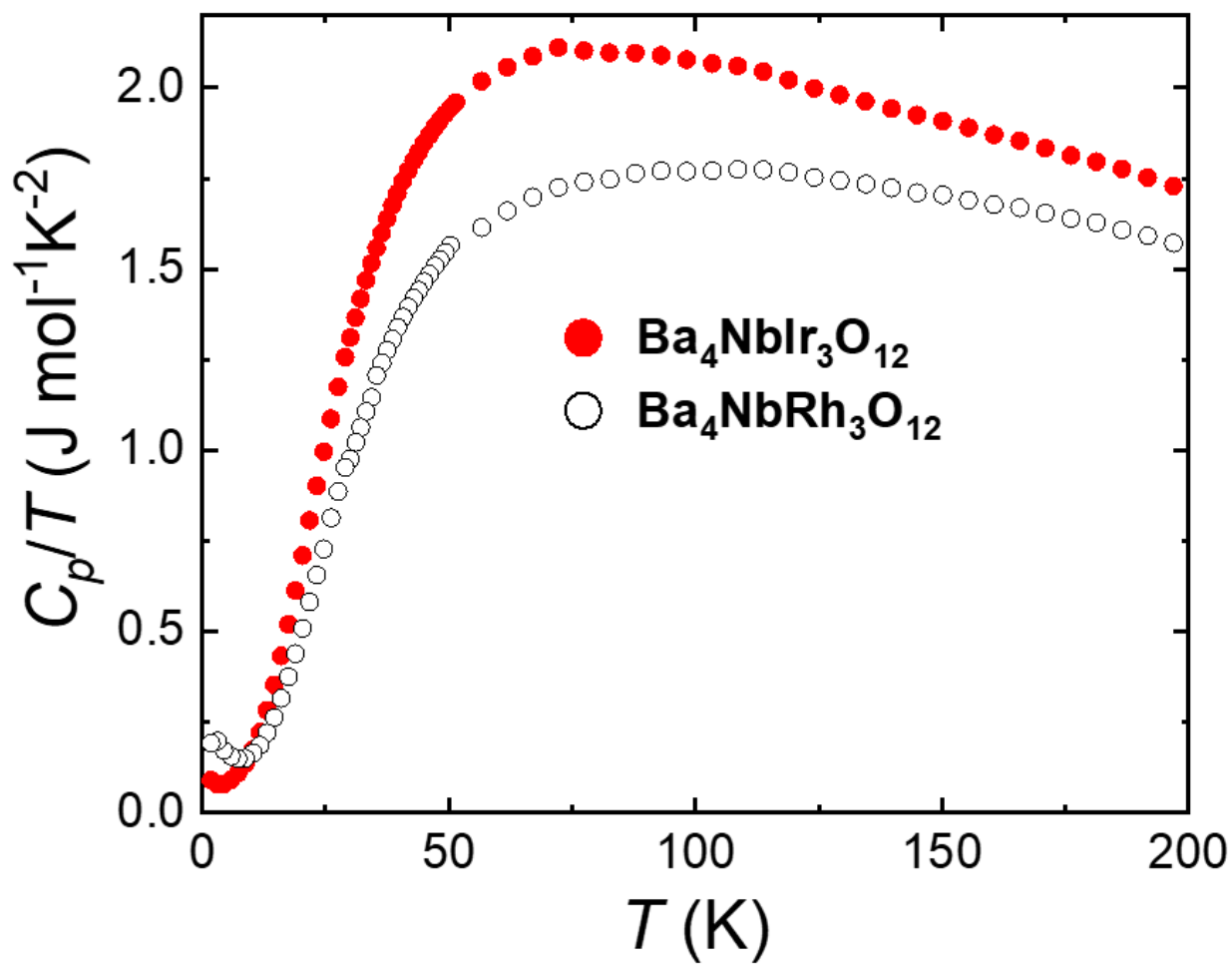

Figure 5: (Color online) Molar heat capacity divided by temperature for Ba$_4$NbIr$_3$O$_{12}$ (red circles) and Ba$_4$NbRh$_3$O$_{12}$ (black open circles) measured from 1.8 K to 200 K. Both materials show features below 10 K (Ba$_4$NbIr$_3$O$_{12}$).

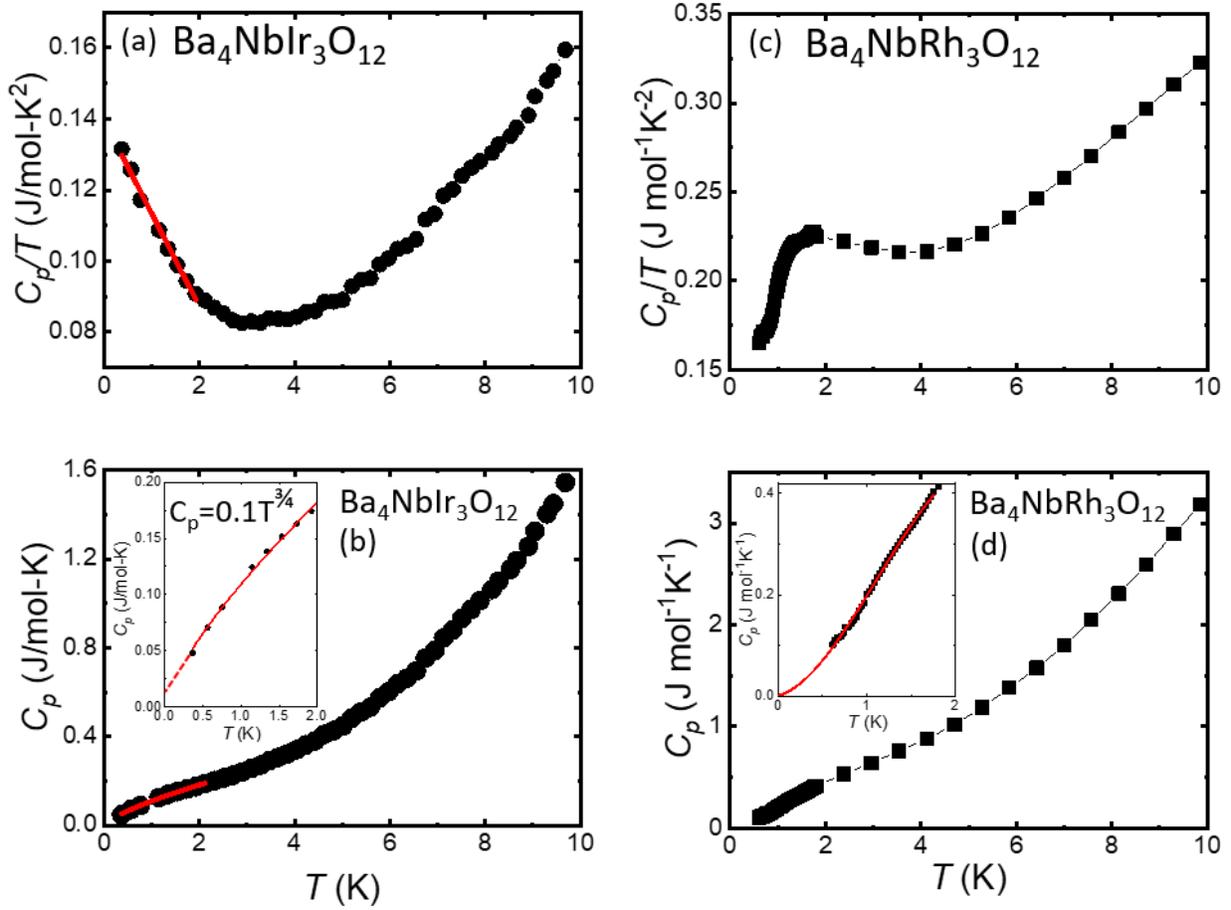

Figure 6: (Color online) (a), (c) – Low temperature molar heat capacity divided by temperature for $Ba_4NbIr_3O_{12}$ and $Ba_4NbRh_3O_{12}$. The linear upturn (red line) at the lowest temperature suggests that $Ba_4NbIr_3O_{12}$ is a candidate spin liquid, while the data for the Rh material appear to be more conventional in character. (b), (d) - Molar heat capacity of $Ba_4NbIr_3O_{12}$ and $Ba_4NbRh_3O_{12}$, respectively, from 10 K to 0.35 K.

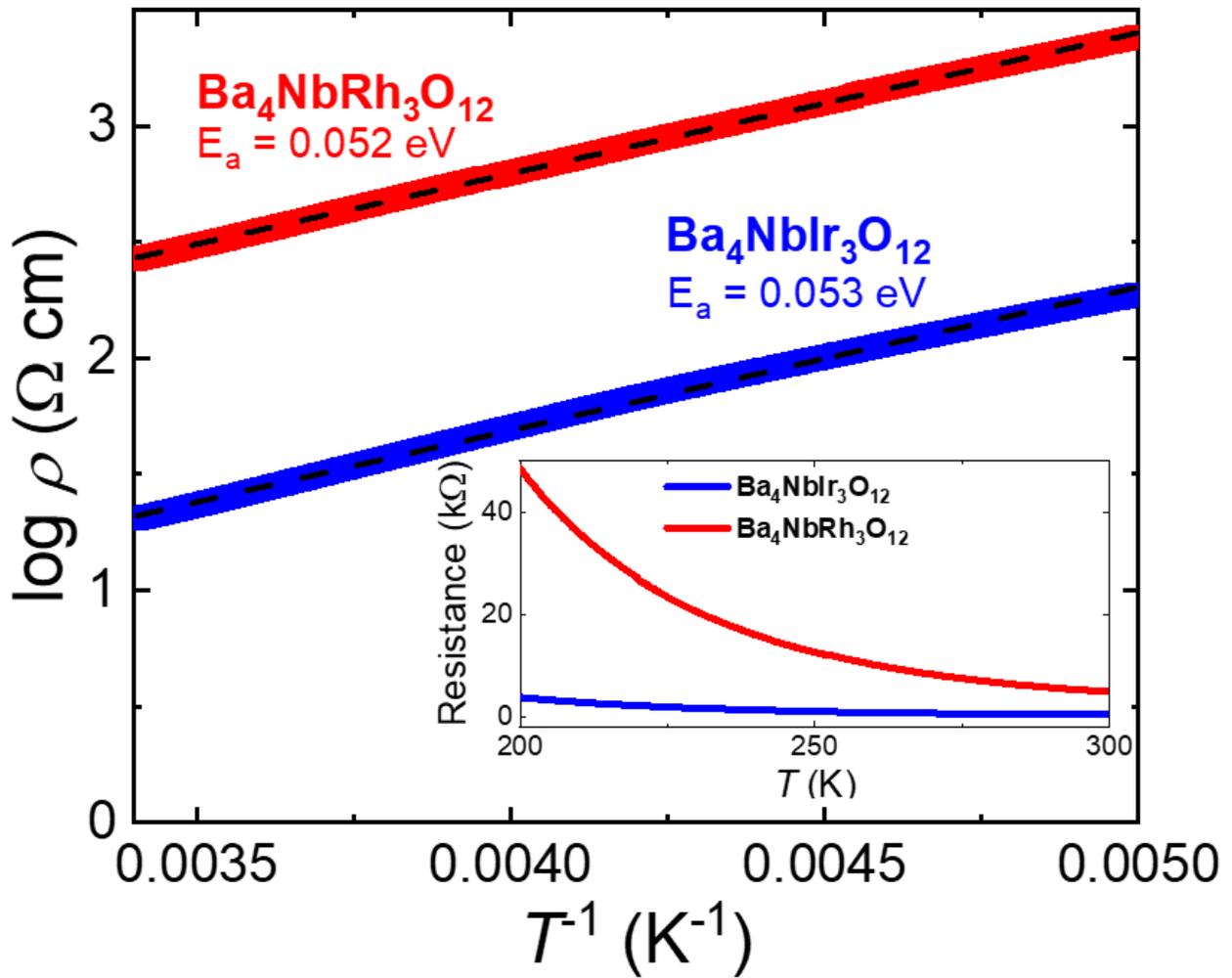

Figure 7: (Color online) Main Panel-The resistivity data in semi-log form was fit to the model $\rho = \rho_o e^{\frac{E_a}{k_b T}}$ (black dashed lines) giving $E_a \approx 0.05$ eV for both $Ba_4NbIr_3O_{12}$ (blue) and $Ba_4NbRh_3O_{12}$ (red). Inset-The raw resistance data for sintered polycrystalline pellets of as a function of temperature.

# Supplementary Information

## The trimer-based spin liquid candidate Ba$_4$NbIr$_3$O$_{12}$

Loi T. Nguyen and R.J. Cava

Department of Chemistry, Princeton University, Princeton, New Jersey 08544, USA

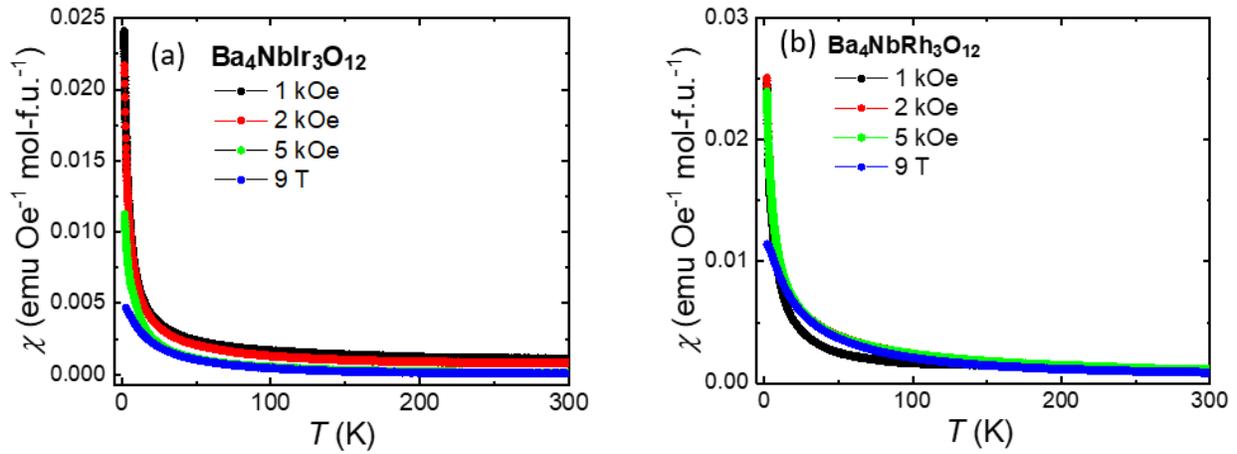

Figure S1: Magnetic susceptibility measured at different fields for (a) Ba$_4$NbIr$_3$O$_{12}$ and (b) Ba$_4$NbRh$_3$O$_{12}$. While the magnetic susceptibility is suppressed by a factor of two in Ba$_4$NbRh$_3$O$_{12}$ from 1 kOe to 9 T, it is five times smaller in Ba$_4$NbIr$_3$O$_{12}$.

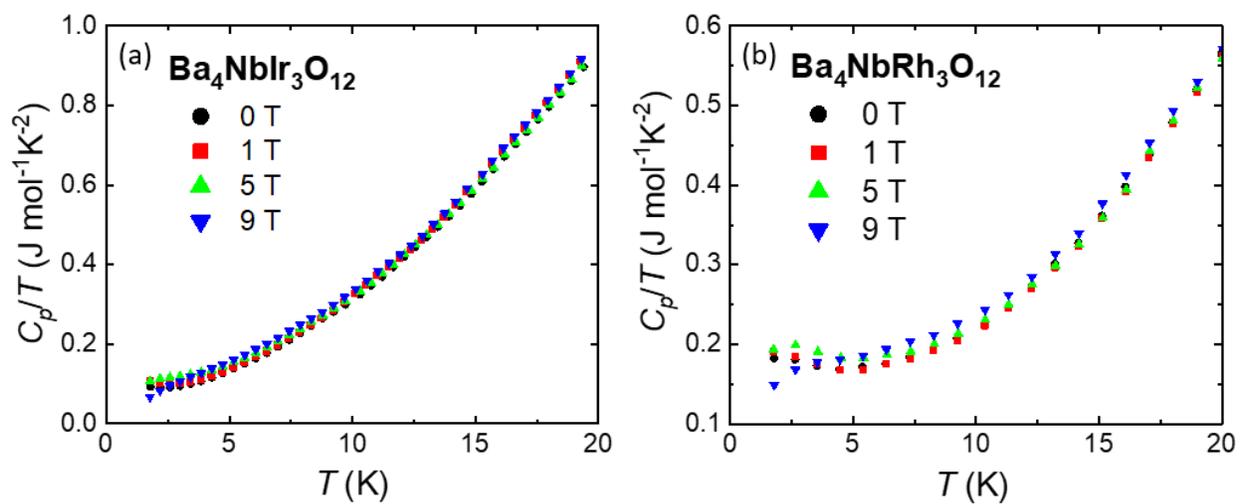

Figure S2: Molar heat capacity divided by temperature measured from 1.8 K to 20 K at 0 T (black circles), 1 T (red squares), 5 T (green up-triangles) and 9 T (blue down-triangles) for (a) $Ba_4NbIr_3O_{12}$ and (b) $Ba_4NbRh_3O_{12}$.